# Ground-state atomic polarization relaxation-time measurement of Rb filled hypocycloidal core-shaped Kagome HC-PCF

T. D. Bradley[1, 2], E. Ilinova[1], J. J. McFerran[1], J. Jouin[3], B. Debord[1], M. Alharbi[1], P. Thomas[3], F. Gerome[1] and F. Benabid[1, 2*]

[1]GPPMM Group, Xlim Research Institute, CNRS UMR 7252, University of Limoges, France
[2]Department of Physics, University of Bath, Claverton Down, BA2 7AY, UK
[3]SPCTS UMR CNRS 7315, Centre Européen de la Céramique, 12 rue Atlantis, 87068 Limoges Cedex, France
*Corresponding author: f.benabid@xlim.fr

We report on the measurement of ground-state atomic polarization relaxation time of Rb vapor confined in five different hypocycloidal core-shape Kagome hollow-core photonic crystal fibers made with uncoated silica glass. We are able to distinguish between wall-collision and transit-time effects in an optical waveguide and deduce the contribution of the atom's dwell time at the core wall surface. In contrast with conventional macroscopic atomic cell configuration, and in agreement with Monte Carlo simulations, the measured relaxation times were found to be at least one order of magnitude longer than the limit set by atom-wall collisional from thermal atoms. This extended relaxation time is explained by the combination of a stronger contribution of the slow atoms in the atomic polarization build-up, and of the relatively significant contribution of dwell time to the relaxation process of the ground state polarization.

Micro-confinement of atomic alkali vapors in hollow waveguides such as hollow core photonic crystal fibers (HC-PCF) has gradually emerged as a timely and an important topic in several applications [1–5]. Such interest is explained by the HC-PCF success in combining compact experimental schemes with enhancement in spectroscopic contrast or optical nonlinearities [6]. This in turn has stimulated efforts in developing HC-PCF designs that overcome some of its intrinsic limitations, which led to the advent of hypocycloidal core-shaped inhibited-coupling (IC) guiding HC-PCF Kagome HC-PCF [7,8]. This hollow waveguide combines ultra-low transmission loss, wide transmission bandwidth, reduced overlap with the silica core surround and larger hollow-core diameters whilst maintaining excellent single mode properties [9–11]. These properties make this fiber an excellent platform for atom optics. In parallel, the atom-light interaction in HC-PCF departs in many ways from conventional configurations and necessitates a fresher look at the spectroscopy measurements of atoms inside the fiber. First, the HC-PCF specific modal properties, such as the guided mode field diameter $d_m$ being very close to the fiber geometrical diameter $d_f$ (typically $d_m \sim 0.7 d_f$), make it difficult to discriminate the optical spectral-line broadening due to transit-time, $\tau_{tt} = (2/\pi) d_m/v$ [12], from that of the atom-wall collisions $\tau_w \approx (V/A)(2\sqrt{\pi})/v$ [13], which reduces to $\tau_w \approx (\sqrt{\pi}/2) d_f/v$ for a cylindrical core-geometry. Here $v, A$ and $V$ are the atoms most probable speed, the fiber hollow-core total-area and volume respectively. This partly explains the measured linewidths found in the early work on electromagnetically induced transparency (EIT) or saturable absorption (SA) in anti-relaxation coated hollow-core waveguides (*e.g.* ARROW's, PBG HC-PCF and IC HC-PCF [1–3]), which do not show clear evidence of anti-relaxation effect. Furthermore, because of the HC-PCF micrometric confining scale and the unusually large surface-to-volume ratio of its core, the atom-wall interaction dynamics can no longer be neglected. First, because of the extremely short $\tau_{tt}$, which remain shorter than 100 ns even for a core diameter as large as 50 µm, the spectral broadening dynamics is often transit-time limited for HC-PCF confined atoms [18], even with low laser power-levels and large frequency detuning. This results in a spectroscopy dominated by only the atoms that are sufficiently slow to complete a transition Rabi cycle before leaving the interaction volume. Second, the close vicinity of the atoms to the large wall surface, their mean dwell time $\tau_{dw}$ at the trapping surface potential [14,15] can no longer be neglected as its range of $\tau_{dw} \sim 80\,ns - 55\mu s$ [16,17] indicates that it can dominate the other dephasing time-scales. This new light-atom hosting configuration thus calls for spectroscopic measurements capable of assessing and discriminating the three mentioned-above time-scales.

Within this context, we report, for the first time to our knowledge, on magneto-optical spectroscopy (MOS) of Rb vapor confined in hypocycloidal core shaped Kagome HC-PCF and assess experimentally and theoretically their ground state polarization relaxation dynamics. The results show that the ground state population relaxation time, $\tau_1$, can be up to ten-fold longer than $\tau_w$ and $\tau_{tt}$. We show that it is dominated by the strong contribution of slow atoms in the Zeeman-optical pumping induced atom polarization build-up and by the larger $\tau_{dw}$ relative to the other relaxation timescales.

Fig. 1(a) schematically illustrates the experimental set-up. One of the five 8 cm long Rb loaded Kagome HC-PCFs placed in a high vacuum chamber is fully Zeeman optically pumped using a circularly polarized laser beam with 650 nW of power (pump beam). The beam is emitted from an extended cavity diode laser (ECDL) whose frequency is set at 500 MHz blue-detuned from the $^{85}$Rb D2 absorption line ($5S_{1/2}F=3\rightarrow5P_{3/2}F'=2, 3, 4$) (Fig. 1(b)). The Zeeman sub-level degeneracy is lifted by applying a dc magnetic field (≤3.5 mT) along the guidance axis of the optical fibers using a solenoid wound around the vacuum chamber. The electron spin randomization rate $\tau_1$ was measured using a variation of Franzen technique [19]. This consists of abruptly shutting off the pump beam with a chopper at 70 Hz frequency rate, and detecting during this pump "dark time", the exponentially decaying atom polarization by recording the polarization rotation, $\varphi$, of a linearly polarized weak probe beam at the fiber output. The probe beam is extracted from the same laser as that of the pump and has a power of 60 nW. The probe polarization rotation is measured by extracting a signal which is proportional to $(P_2 - P_1)/(P_2 + P_1)$ from a polarimeter based on balanced photodetectors. Here, $P_1$ and $P_2$ are the probe-beam power of the two orthogonal polarization components transmitted through the HC-PCF and split by a polarizing beam-splitter. Examples of recorded signal traces for the case of two different kagome HC-PCFs from the fiber set are shown in Fig. 1(c). Each trace is then fitted to a single exponential function, from which the probe polarization rotation time constant $\tau_\varphi$ is extracted. The latter is, in our condition, directly related to the electronic spin relaxation time $\tau_1$ by $\tau_\varphi^{-1} \approx \tau_1^{-1}/18$ [19]. The factor 18 is the calculated ratio between the state-change probability in the electron spin and the atom total spin after a single collision [20]. Furthermore, because of the low atomic density we operate at ($\sim n < 10^{10} cm^{-3}$) the Rb-Rb spin exchange collisional rate $\gamma_{ex} = 1/\tau_{ex}$ ($\tau_{ex} = 1/(n\sigma_{ex}v) \approx 180 ms$) can be neglected. Hence, the rate of electron spin randomization is given by $\gamma_{er} \sim 1/\tau_w$, and the atomic polarization relaxation time is reduced to $\tau_1^{-1} \approx \tau_w^{-1} + \tau_{dw}^{-1}$. Consequently, this experimental

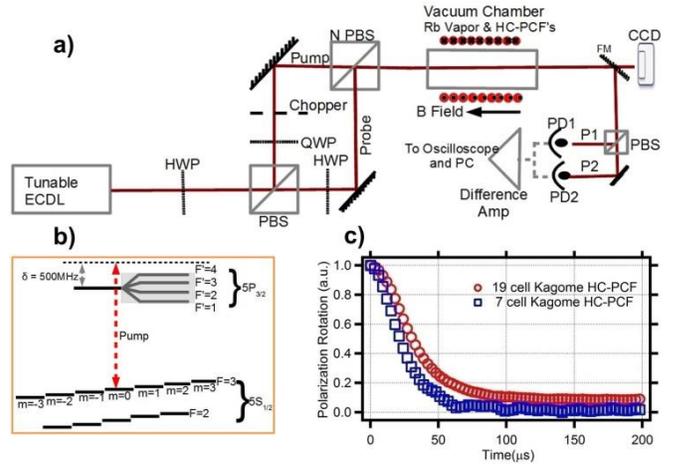

FIG. 1 a) Experimental setup for probing atomic polarization state of Rb vapour loaded in HC-PCFs. FM: flip-mirror; PD: photodetector; PBS: polarizing beam-splitter, NPBS: non-polarizing beam-splitter. $P_1$ and $P_2$ are the laser power received by PD1 and PD2 respectively. HWP: half waveplate, and QWP: quarter waveplate b) $^{85}$Rb energy level structure in the presence of the DC magnetic field. c) Typical signal traces recorded for two different HC-PCFs during the pump dark time.

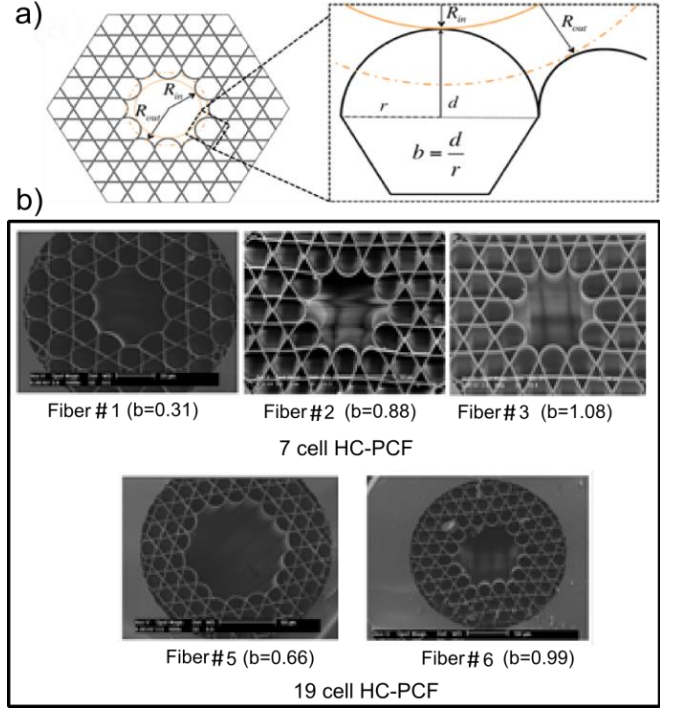

FIG. 2 a) Schematic representation of hypocycloid core shape Kagome HC-PCF (*lhs*), and parameterisation of the hypocycloid curvature (*rhs*). b) Scanning electron micrographs of the five explored Kagome HC-PCF with their respective *b* values.

protocol applied to several HC-PCFs with different core radii can directly measure the HC-PCF atom-wall collision dynamics induced population decay and the dwell time.

The fiber set consists of five hypocycloidal-like core-shaped Kagome HC-PCF's [7,8,20,21] with different contour curvature values and radii. Fig. 2 summarizes the physical properties of the fibers. The core-contour geometry is fully parameterized by the inner radius $R_{in}$, the cup curvature parameter $b$, and the ratio between the inner radius and the outer radius $\theta = R_{out}/R_{in}$ (Fig. 2a). The parameter $b$ is defined as the ratio of the distance from the outer edge of the arc to the half-length of the chord transecting the largest diameter of the cup [10]. It is noteworthy that the larger $b$ is the better the fiber transmission performance is and the closer the fiber modal content is to a single mode guidance regime [10]. Three of the HC-PCF's have a 7-cell core defect with inner core radius $R_{in}$ ranging between ~15 μm and ~30 μm, and outer radius of $R_{out}$ ranging from ~16 μm and ~39 μm (see top of Fig. 2(b) and Table 1). The remaining 2 fibers are 19-cell core defects with core inner radii of 38 μm and 48 μm (rhs of Fig. 2(b)). All the fibers have been fabricated for low loss guidance around 780nm with a range of 70 dB/km-300 dB/km. The fibers are loaded with natural Rb vapor for a period of 3.5 weeks. At the end of the vapor loading process the current source to the Rb getter is switched off to minimize the influence of the Rb source on the relaxation time measurement [19]. During the data collection, the Rb vapor density was measured and was found to remain steady over the corresponding time period.

Table 1) $R_{in}, R_{out}, b$ are fiber physical dimensions, $n$ is the Rb atomic densities. $\tau_{tt}, \tau_w$ and $\tau_1$ are the transit time, atom-wall collision relaxation time, and measured relaxation time respectively.

| Fiber | $R_{in}/R_{out}$ (μm) | b | n ($10^9$ cm$^{-3}$) | $\tau_{tt}$ (μs) | $\tau_w$ (μs) | $\tau_1$ (μs) |
|---|---|---|---|---|---|---|
| #1 | 15/15.8 | 0.31 | 1.0 | 0.05 | 0.10 | 0.97 |
| #2 | 25.5/30.6 | 0.88 | 5.0 | 0.08 | 0.17 | 1.07 |
| #3 | 29.5/39.3 | 1.08 | 7.0 | 0.10 | 0.19 | 1.06 |
| #4 | 38/42 | 0.66 | 6.5 | 0.13 | 0.25 | 1.045 |
| #5 | 48/51 | 0.99 | 6.7 | 0.16 | 0.31 | 1.18 |

Table 1 lists the measured atomic density $n$ along with the inner and outer radius, the $b$ parameter, the transit time $\tau_{tt}$, the wall-atom collision time $\tau_w$, and the measured ground-state atomic polarization relaxation time $\tau_1$. Because of the non-circular geometry of the core-contour, $\tau_w$ takes a different expression from the circular contour and is given in the supplementary materials (SM)).

The salient and common feature of the measured $\tau_1$ is that they are all more than one order of magnitude larger than the atom-wall relaxation time. These longer $\tau_1$ values occur because of the combination of the enhanced initial transit time of polarized atoms and the effect of the dwell time of the atoms in the wall trapping potential. The increased transit-time results from the slow polarization build-up rate at our given experimental conditions. This is in turn means that only the transversally slow atoms are polarized. This is explained in Fig. 3, which summarizes the atomic polarization build-up dynamics as deduced from solving the optical Bloch equations. Fig. 3(a) shows the seven Zeeman ground-state population evolutions when the atoms are excited by the pump at our experimental conditions. The atomic population is fully optically-pumped to the ground state Zeeman sublevel $|F = 3, m = +3\rangle$ after a certain build-up time $\tau_{bp}$, following a time evolution $\propto \left(1 - e^{-\frac{t}{\tau_{bp}}}\right)$, with $\tau_{bp}$ found to be approximately 2ms. This is more than 3 orders of magnitude larger than transit-time limit $\tau_{tt}$. It is noteworthy that the buildup time $\tau_{bp}$ can be further increased either by increasing the detuning from the resonance or decreasing the pump intensity ($1/\tau_{bp} \sim I/\delta^2$ [24]), as numerically calculated and shown in Figs. 3(b) and 3(c). In our experiment, the blue detuning of 500 MHz from F=3 to F'=2,3,4 line is set to optimize the polarization rotation amplitude, and the pump power value (coupled intensity in the range of 10-100 mW/cm²) is set to produce a sufficient signal to noise ratio.

As a consequence of such a long polarization build-up time, only atoms whose time-of-flight through the beam without depolarization is longer than $T_{bp}$ significantly contribute to the ground-state atomic polarization (i.e. populating $|F = 3, m = +3\rangle$) [18]. The velocity-distribution $f(v)$ of these polarized atoms is illustrated by the red curve in Fig. 3(d) along with the Maxwellian distribution (blue filled curve) for comparison (See SM for details). The results readily show that the polarized atom velocity distribution significantly deviates from the Boltzmann-Maxwell thermal distribution, and peaks at much smaller velocity subclasses. The calculations for the fiber #1 experimental conditions show that the atom polarization most probable speed was found to be $v_{pol} \approx 8 \, m/s$, which is ~20 times slower than the most probable speed $v$ for thermal distribution. Equivalently, when the pump laser is turned off, the polarized atoms before their depolarization and thermalization by wall-collision cross the interaction volume with a transit time 20 slower than the one experienced by thermal atoms. In order to assess the impact of such a nonthermal velocity distribution on the atom-wall relaxation time, we carried out a Monte Carlo (with $10^5$ realizations) based calculation to trace the trajectory of an initially polarized atom during the dark time (See SM). Here, for each atom trajectory we assume that each collision corresponds to the single occurrence of spontaneous electron spin state-change (electron spin randomization) and that the nucleus spin momentum projection does not

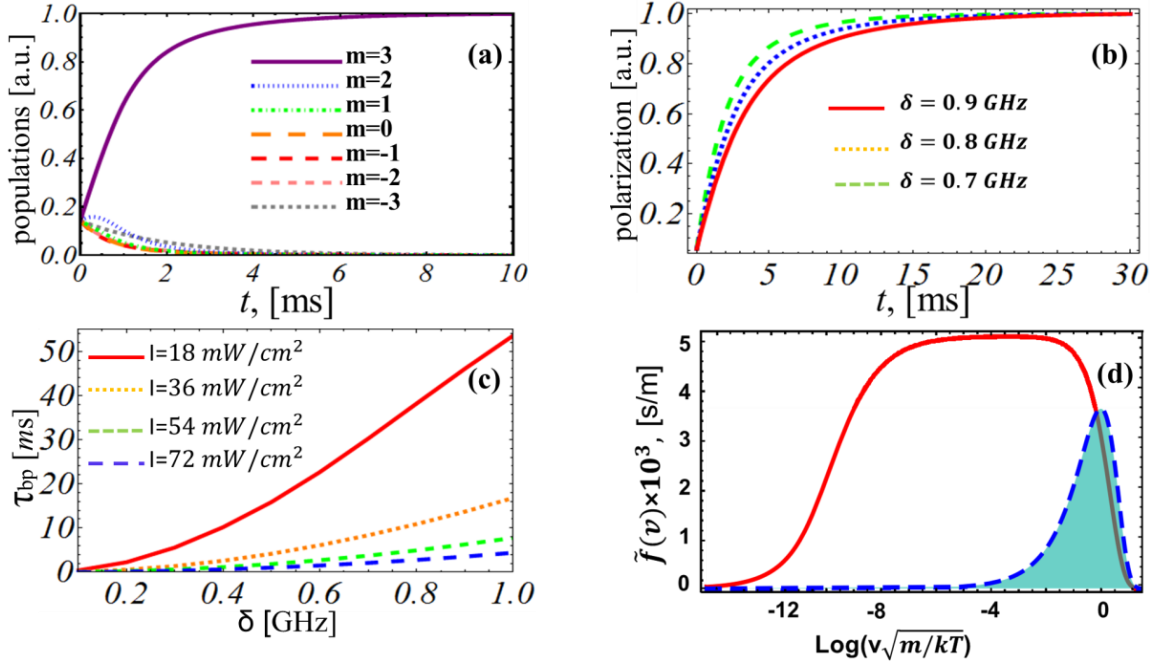

FIG. 3) (a) F=3 Zeeman sublevels normalized-population evolution in the presence of a pumping laser and in the experimental condition with fiber #1. (b) Atom polarization time evolution for pump laser intensity of 18 mW/cm² and detuning of 700 MHz (solid red curve), 800 MHz (dotted orange curve) and 900 MHz (dashed green curve). (c) Atom polarization build-up time versus the pump frequency detuning for the intensity of 18 mW/cm² (solid red curve), 36 mW/cm² (dashed orange curve), 54 mW/cm² (dashed green curve), 72 mW/cm² (dashed blue curve). (d) The effective transverse speed probability density function for the polarized atoms (red curve). For comparison the Rb thermal Maxwell-Boltzmann distribution is shown (blue filled curve).

change during the collision [22][23]. This is implemented in the model, by associating the atom polarization with a pseudo-spin $S$ that can take a value of 1 or 0 during each realization. The pseudo-spin value S=1 corresponds to polarized atom (i.e. prepared in a given superposition of Zeeman states) [20] with the probability of 1/18 to drop to 0 after each single collision. We also assume that initially we have S=1, and that the atomic velocity-vector stochastically changes after each collision with the wall, and its newly acquired absolute value obeying the Maxwell-Boltzmann distribution. This means that in the absence of pump laser the atom ensemble relaxes to a thermal distribution between Zeeman sublevels and its polarization decays to zero with a probability of 1/18 upon every wall-collision. Finally, the time dependence of the average polarization in one ensemble, the lifetimes of the polarization in the volume and the effective polarization in the interaction volume were deduced from the results of the $10^5$ realizations. The interaction volume should be understood as follows. The atoms interact with the probe beam only when they are in the laser effective area (shown in yellow on fig. 4 (b)), where the mode-field radius is taken to be $0.7R_{in}$ [10]. Finally, each collision is associated with a dwell time $\tau_{dw}$ [24]. The role of the dwell-time in the polarization signal decay process can be understood in analogy with the extension of the effective fiber core inner diameter while assuming the collisions with the wall being instantaneous. Consequently, one can deduct from each experimentally measured $\tau_1$ a trial value for the dwell time of $\tilde{\tau}_{dw}$. This is used in the Monte-Carlo model as follows: the dwell-time value during each collision and for a given realisation is taken to be random over a range of 0 to $2\tilde{\tau}_{dw}$. However, the average of dwell-time, $\tau_{dw}$, over all the realisations is taken as a unique fitting parameter to the measured $\tau_1$ obtained with the five fiber configurations. Finally, because of our off-resonance optical pumping scheme and the very low atomic density we neglect the radiation trapping, which is another source of polarization relaxation [25].

Results from the Monte-Carlo simulations are summarized with the experimental results in Fig 4. Figure 4(a) shows an excellent agreement between the measured and the calculated relaxation time $\tau_1$. For comparison, the calculated $\tau_w$ and $\tau_{tt}$ are also shown, and one clearly notices the significantly longer polarization relaxation time $\tau_1$ than the limit set by the atom-wall collision rate of thermal Rb. The results give a fitted $\tau_{dw}$ of 0.7µs, which is consistent with the measured one reported by Zhao et al. [14]

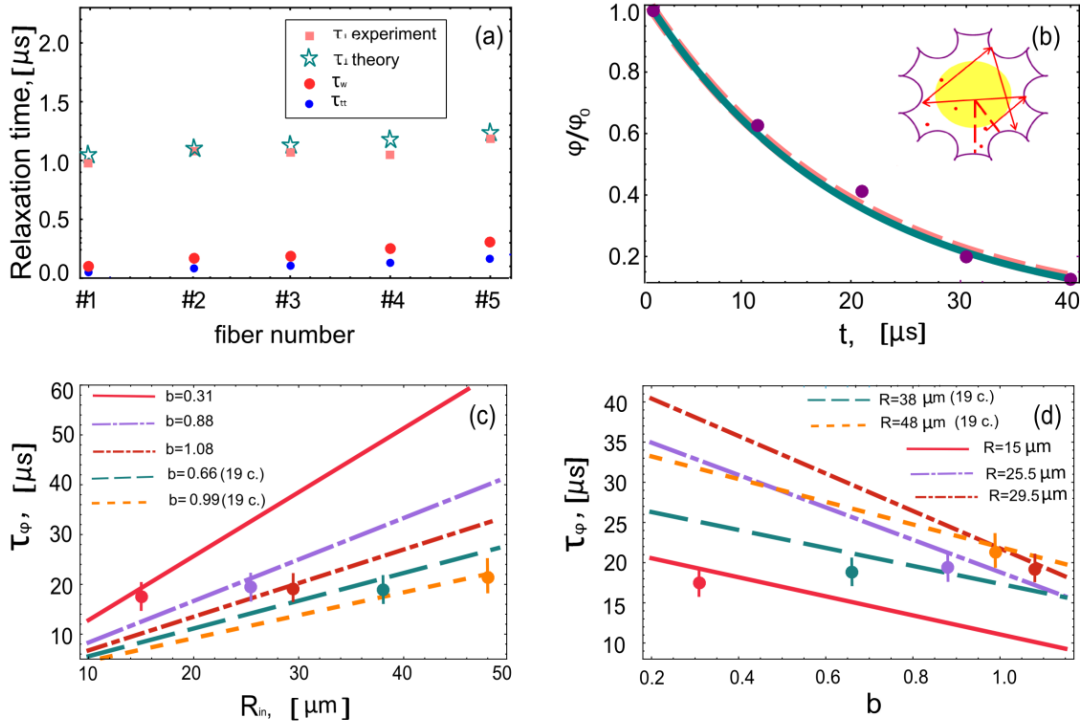

FIG. 4) (a) Theoretical and measured $\tau_1$, atom-wall relaxation time and transit-time for the tested fibers. (b) Time evolution of the atom polarization induced optical rotation as calculated (pink dashed-line), measured (purple dots) and its single exponential fit (green solid line). Inset: Schematic atom-trajectory inside the HC-PCF. The yellow disk represents the beam effective area. (c) Relaxation time as a function of radius at fixed curvature parameter: (7 cell) $b$ =0.31 (red curve), $b$=0.88 (purple curve), $b$ =1.08 (red dashed curve), (19 cell) $b$ =0.66 (orange curve), $b$ =0.99 (green curve), (d) Relaxation time as a function of curvature at fixed inner radius: (7 cell) R=15 (red curve), R=25.5 (purple curve), R=29.5 (red dashed curve), (19 cell) R=38 (green curve), R=48 (orange curve). Circles represent the experimental points.

for coated Pyrex glass surface. The very good agreement between experiment and simulation is further illustrated in Fig. 4(b), which shows the atom polarization time-evolution in the dark for the case of fiber # 1 as measured (solid points) and Monte Carlo calculated (red dashed curve). The figure also shows the single-time exponential decay fit to the experiment (solid green curve). To complete the study, Fig. 4(c) and 4(d) show how the polarization decay scales with the fiber geometrical parameters. We readily observe that at fixed curvature (Fig. 4(c)), the relaxation time increases linearly with radius as expected from conventional macroscopic cylindrical vapor cells [26]. On the other hand, the relaxation time dependence with curvature (Fig. 4(d)) shows a decrease of the relaxation time with increasing $b$. This may be expected since an increase in $b$ implies an increase in the surface-to-volume ratio. This behavior is in quantitative agreement with the experimental data for the five different fibers.

In conclusion we have demonstrated using a combination of experimental protocol and Monte-Carlo simulation the major dephasing time scales in the ground state of HC-PCF confined Rb. Unlike in conventional cells, the atom polarization dynamics is no longer dominated by the transit-time or by atom-wall collisions but by the stronger contribution of the slow atoms in the atom polarization build-up and to the dwell time of the atom at the core wall surface.

# Supplementary materials for the manuscript "Ground-state atomic polarization relaxation-time measurement of Rb filled hypocycloidal core-shaped Kagome HC-PCF"

T. D. Bradley[1,2], E. Ilinova[1], J. J. McFerran[1], J. Jouin[3],
B. Debord[1], M. Alharbi[1], P. Thomas[3], F. Gerome[1], and F. Benabid[1,2]

1 *GPPMM Group, Xlim Research Institute, CNRS UMR 7252, University of Limoges, France*
2 *Department of Physics, University of Bath, Claverton Down, BA2 7AY, UK and*
3 *SPCTS UMR CNRS 7315, Centre Europen de la Cramique, 12 rue Atlantis, 87068 Limoges Cedex, France*

In this supplementary materials we give some further details on theoretical model used in the main manuscript.

## 1. Calculating the wall collision rate for the 7 and 19 cell geometry fibers

The rate of the atom wall collisions is determined as $\gamma_{col} = \frac{A}{V}$, where A is the collisional surface and V is the volume of the atomic container and vt is the transverse speed of the atoms. To determine the ratio A/V for the 7 and 19 cell Kagome fibers we used the approximate expression for the perimeter P of the half ellipse describing the shape of the of the inner core cups, $P(r,\eta) \sim \pi\sqrt{a^2+b^2}$
where a and b are the half axes of the ellipse. The ratio of collision area A/V to the given geometries are given then by

$$(A/V)_{7cell} = 4\sqrt{2}\pi(\eta_2 - 2\sqrt{3})\frac{\left(\sqrt{(b^2-4\sqrt{3}b+12)\theta^2+7(b-2\sqrt{3})\theta+13}+\sqrt{b^2+1}\right)}{R_{in}(\pi(2(\sqrt{3}b-6)\theta+4b+7\sqrt{3})-30\sqrt{3})}$$

$$(A/V)_{19cell} = 4\sqrt{2}\pi(b^2 - 6\sqrt{3}b + 28)\frac{\left(\sqrt{(b^2-6\sqrt{3}b+28)\theta^2-11\sqrt{b^2-6\sqrt{3}b+28}\theta+31}+2\sqrt{b^2+1}\right)}{R_{in}((2\sqrt{3}\pi\theta+16)\sqrt{b^2-6\sqrt{3}b+28}+(2-\pi)(8b+11\sqrt{3}))}$$

where θ=$R_{out}/R_{in}$ is the ratio of the outer and inner core radiuses of the fiber as it is shown on Fig 2 of the main text, and b is the ratio between the lengths of the axes of half ellipse approximating the shape of the bigger cup.

### B. Initial velocity distribution of polarized atoms

The velocity selective optical pumping leads to the non Maxwellian initial transverse velocity distribution of polarized atoms. This can be understood as follows. The dependence of the atomic polarization on the time tpump spent inside the pump laser beam is given by
$$P(t) = 1 - e^{-t/T_{bp}}$$
where the parameter Tpb is the characteristic polarization build up time which depends on the de-tuning and the intensity of the pump laser. The velocity dependence of the atomic polarization created during a single pass of the atom through the pump beam cab be obtained then as

$$P_{sp}(v) = 1 - e^{-\sqrt{2}R_{int}/(vT_{bp})}$$
where $\sqrt{2}$Rint is the average length of the path through the interaction cross section.

After a single collision with the wall the atom can keep its total momentum projection unchanged. Consequently the pumping can occur during several passes of the atom through the interaction volume. To calculate the resulting velocity dependence of the atomic polarization P(v), one needs to average over the all possible trajectories of the atom during the pump process, taking to account the probability of depolarization and the velocity change after each collision with the wall. At the condition when the pumping time is much longer than the typical polarization build up time, the final atomic polarization velocity dependence can be parameterized as:

$$P(v) = p_0 - (1-p_0)(1 - e^{-\frac{\sqrt{2}R_{int}}{vT_{bp}}})$$

where p0 is constant. One can show that when the polarization build up time is much larger then the time between the two following collisions with the wall, so that Tbp >> ncTw, where nc is the average number of collisions experienced by the atom before it depolarizes, the value of p0 is much less than 1 and one can approximate P(v) as

$$P(v) \approx 1 - e^{-\frac{\sqrt{2}R_{int}}{vT_{bp}}}$$

The effective transverse velocity distribution of polarized atoms can be introduced then as

$$\tilde{f}(v) = \frac{1}{N}f(v)(1 - e^{-\frac{\sqrt{2}R_{int}}{vT_{bp}}})$$

where

$$N = \int_0^\infty f(v)(1 - e^{-\frac{\sqrt{2}R_{int}}{vT_{bp}}})dv$$

The slowing factor fs can be introduced as the ratio of most probable velocities of sub ensemble of polarized atoms and the whole ensemble of thermal atoms: fS=$\tilde{v}_p \cdot \sqrt{2m/kT}$, where $\tilde{v}_p$ is the solution of the equation

$$T_{bp}v(2v^2 - v_p^2)\left(e^{\frac{R}{vT_{bp}}} - 1\right) + Rv_p^2 = 0$$

where vp=$\sqrt{2m/kT}$, R is the laser beam radius. For Tbp=1ms one has $\tilde{v}_p \sim 8\ m/s$. The ratio vp/$\tilde{v}_p$=169/8=21.
To find the average transverse velocity of the polarized atoms, one needs to calculate the value:

$$\overline{v_{pol}} = \left(\int_0^\infty f(v)(1-e^{-R\sqrt{2}/(vT_{bp})})v\,dv\right) / \left(\int_0^\infty f(v)(1-e^{-R\sqrt{2}/(vT_{bp})})\,dv\right)$$

**C. Monte-Carlo simulations**

To calculate the time dependence of the polarization rotation signal we calculate the evolution of the number of polarized atoms inside the interaction volume.

It's worth noting that the number of polarized atoms inside the interaction volume Nint (t) (at given moment of time) is different than in the whole volume. To calculate Nint(t)/ Nint (0) we trace the trajectories of Nint (0) =100000 initially polarized atoms. We associate the polarization (total momentum projection state) with the pseudo spin S, which can be 1 or 0. We say that at the moment of time t=0 there are Nint (0) atoms which have the pseudo spin value equal to 1.

After each collision with the wall the atom can change its pseudo spin value to zero with the probability 1/18 (we assume that each collision completely randomize the electron spin of the atom). After the pseudo spin of the atom becomes zero it remains zero until the end of the probe period. The number Nint (t)=$\sum_{i=1}^{N_{int}(0)} p_i(t)$, where $p_i = 1$, if $S_i = 1$ and the atom is inside the interaction volume, otherwise $p_i = 0$.

We consider the atom (at the moment of the pump switch off) can be at the arbitrary position in the fiber, have the random direction and module of the transverse velocity. Taking to account the symmetry of the fiber its enough to consider the dynamics of the atoms which are initially in the sector marked red on the the Fig. 1 below. Each time the atom collides with the wall it changes it's velocity randomly (according to Maxwell distribution).

The scope of this paper does not include calculating the average dwell time $T_{dw}$ which the atom spends near the wall surface after each collision (without being depolarized) and we take it as unknown parameter. We perform the Monte-Carlo simulations for a set of values of $T_{dw}$ in the interval from 0 to 2μs.

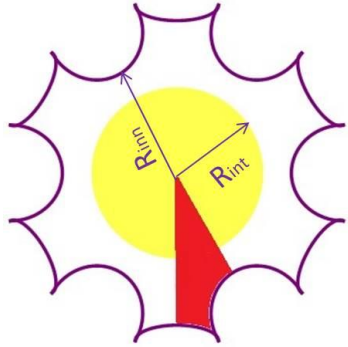

During the MC simulations of atomic spin dynamics (at given fiber geometry and fixed value of the average dwell time $T_{dw}$), for each collision with the wall we take the current value the dwell time $t_{dw}$ randomly from the interval 0 <tdw< 2 $T_{dw}$, assuming that it is uniformly distributed on this interval.

Based on the values of T1 calculated for different fixed values of $T_{dw}$ for each fiber geometry we determine the dependence T1($T_{dw}$). After that we determine the optimal value of $T_{dw}$ corresponding to the least discrepancy of the calculated and experimental values of T1 considering all set of 6 fiber geometries.